\documentclass[aps,pra,twocolumn,superscriptaddress]{revtex4}
\usepackage{graphicx}
\usepackage{amsmath}

\begin{document}

\title{Two-photon Interference with Non-identical Photons}

\author{Jianbin Liu}
\email[]{liujianbin@mail.xjtu.edu.cn}
\affiliation{Electronic Materials Research Laboratory, Key Laboratory of the Ministry of Education \& International Center for Dielectric Research, Xi'an Jiaotong University, Xi'an 710049, China}

\author{Yu Zhou}
\affiliation{MOE Key Laboratory for Nonequilibrium Synthesis and Modulation of Condensed Matter, Department of Applied Physics, Xi'an Jiaotong University, Xi'an 710049, China}

\author{Huaibin Zheng}
\affiliation{Electronic Materials Research Laboratory, Key Laboratory of the Ministry of Education \& International Center for Dielectric Research, Xi'an Jiaotong University, Xi'an 710049, China}
\affiliation{MOE Key Laboratory for Nonequilibrium Synthesis and Modulation of Condensed Matter, Department of Applied Physics, Xi'an Jiaotong University, Xi'an 710049, China}

\author{Hui Chen}
\affiliation{Electronic Materials Research Laboratory, Key Laboratory of the Ministry of Education \& International Center for Dielectric Research, Xi'an Jiaotong University, Xi'an 710049, China}

\author{Fu-Li Li}
\affiliation{MOE Key Laboratory for Nonequilibrium Synthesis and Modulation of Condensed Matter, Department of Applied Physics, Xi'an Jiaotong University, Xi'an 710049, China}

\author{Zhuo Xu}
\affiliation{Electronic Materials Research Laboratory, Key Laboratory of the Ministry of Education \& International Center for Dielectric Research, Xi'an Jiaotong University, Xi'an 710049, China}

\begin{abstract}
The indistinguishability of non-identical photons is dependent on detection system in quantum physics. If two photons with different wavelengths are indistinguishable for a detection system, there can be two-photon interference when these two photons are incident to two input ports of a Hong-Ou-Mandel interferometer, respectively. The reason why two-photon interference phenomena are different for classical and nonclassical light is not due to interference, but due to the properties of light and detection system. These conclusions are helpful to understand the physics and applications of two-photon interference.
\end{abstract}


\date{\today}

\maketitle

Two-photon interference is essential to test the foundations of quantum mechanics \cite{epr,bell}, which has been studied extensively \cite{mandel-book,scully-book} since the observation of the first two-photon interference phenomenon \cite{HBT}. In the existed studies, quantum interference is usually employed to describe multi-photon interference with nonclassical light \cite{scully-book,pan-RMP}. It may mislead people to think that the reason why multi-photon interference phenomena with classical and nonclassical light are different is the interferences with classical and nonclassical light are different. In fact, the interferences with photons in classical and nonclassical light are the same, which is the superposition of probability amplitudes for different yet indistinguishable alternatives \cite{feynman-l}. The reason why multi-photon interference phenomena with classical and nonclassical light are different is the properties of light and detection system, not interference.

The superposition principle in quantum physics is not only valid for all orders of interference with photons in classical and nonclassical light, but also valid for interference with massive particles \cite{feynman-l,feynman-p}. It is based on the indistinguishability of different alternatives, which is related to the indistinguishability of particles. It is well-known that identical particles can be indistinguishable. Can non-identical particles be indistinguishable? The answer is dependent on the properties of particles and detection system. Non-identical particles can be indistinguishable for some detection systems, while be distinguishable for other detection systems. In this letter, we will employ photons to show in what conditions there is two-particle interference with non-identical particles. All the second-order interference of light such as two-photon bunching of thermal light \cite{HBT}, Hong-Ou-Mandel (HOM) dip \cite{HOM}, ghost imaging with both entangled photon pairs and thermal light \cite{pittman-1995,valencia-2005}, quantum lithography \cite{boto-2000} can be interpreted by two-photon interference based on superposition principle \cite{glauber-1963, shih-book}. Although classical theory can be employed to interpret the second-order interference of classical light \cite{glauber-1963,sudarshan-1963}, we will employ two-photon interference in Feynman's path integral theory for it is easy to understand the physics of calculations \cite{feynman-l,feynman-p} and will give a unified interpretation for the second-order interference with classical and nonclassical light.

\begin{figure}[htb]
    \centering
    \includegraphics[width=40mm]{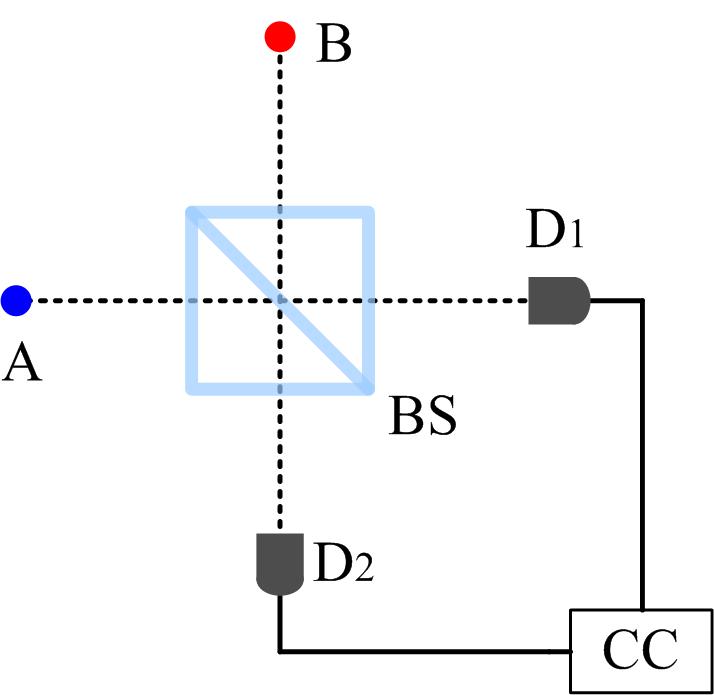}
    \caption{(Color online) Two photons in a HOM interferometer. BS: 1:1 nonpolarized beam splitter. D: Single-photon detector. CC: two-photon detection system.}\label{two}
\end{figure}

We will take two photons with different wavelengths in a HOM interferometer shown in Fig. \ref{two} for example. The wavelengths of photons A and B are different and they incident to two input ports of a 1:1 nonpolarized beam splitter (BS), respectively. There are two different alternatives to trigger a two-photon coincidence count, which are $A\rightarrow D_1, B\rightarrow D_2$ and $A\rightarrow D_2, B\rightarrow D_1$, respectively. $A\rightarrow D_1$ is short for photon A goes to detector 1 (D$_1$) and other symbols are defined similarly. If photons A and B are identical, these two different alternatives are indistinguishable and there is two-photon interference \cite{kaltenbaek-2006}. Two-photon interference is still observed with photons of different bandwidths \cite{bennett-2009} and different central frequencies \cite{kaltenbaek-2009,legero-2004,liu-2014}, which indicates that photons of different spectrums can be indistinguishable.

Two-photon interference with non-identical photons can be understood in quantum physics by taking measurement into consideration. Photon is usually detected by photoelectric effect in single-photon detector. It has been proved by Forrester \textit{et al.} that the time delay between photon absorption and electron release is significantly less than $10^{-10}$ s \cite{forrester-1955}. Hence in principle, the accuracy of time measurement with photoelectric effect can be significantly less than  $10^{-10}$ s \cite{von}.  Heisenberg's uncertainty principle with energy and time measurement uncertainties is $\Delta E \Delta t \geq \hbar$ \cite{bohm-QM}. With $E=\hbar \omega$ for photons, it is easy to get $\Delta \omega \Delta t \geq 1$, where $\Delta \omega$ and $\Delta t$ are the frequency and time measurement uncertainties, respectively.  For a detector that can response to $\omega_A$ and $\omega_B$, these two photons are indistinguishable if $|\omega_A-\omega_B|$ is less than $1/\Delta t$. It is the reason why there is two-photon interference with photons of different spectrums in Refs. \cite{bennett-2009,legero-2004,liu-2014}.

\begin{figure}[htb]
    \centering
    \includegraphics[width=60mm]{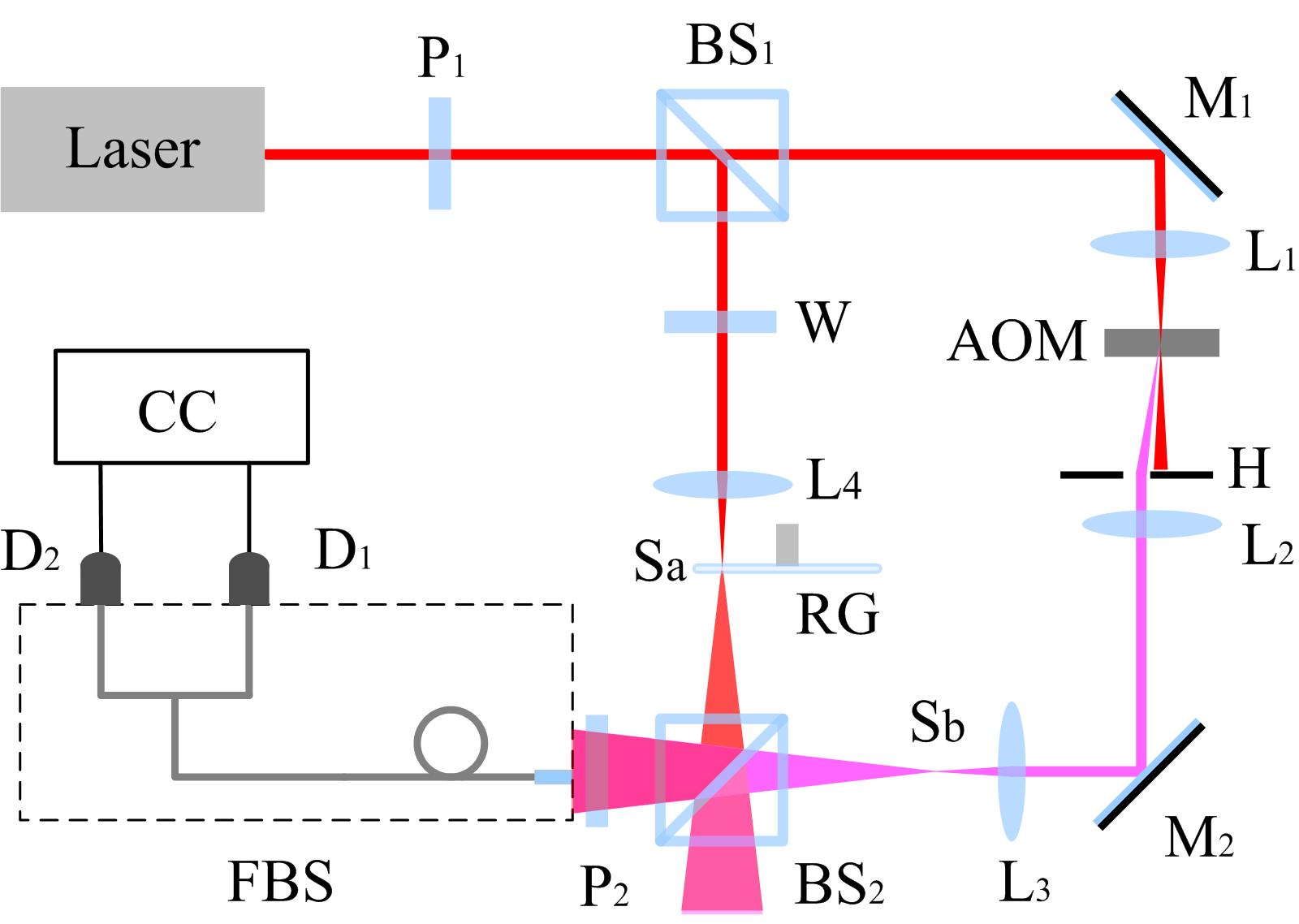}
    \caption{(Color online) The experimental setup for two-photon interference. Laser: 780 nm single-mode laser with bandwidth of 200 kHz. P: Polarizer. BS: 1:1 nonpolarized beam splitter. W: $\lambda/2$ wave plate. RG: Rotating ground glass. S: Light source. L: Lens. M: Mirror. AOM: Acoustooptic modulator. H: Pinhole. FBS: Fiber beam splitter. D: Single-photon detector. CC: two-photon coincidence count detection system. See text for details.}\label{setup}
\end{figure}

The scheme shown in Fig. \ref{setup} is employed to test the predictions. The laser is a single-mode continuous wave laser with 780 nm central wavelength and 200 kHz frequency bandwidth. P$_1$ and P$_2$ are polarizers. BS$_1$ and BS$_2$ are 1:1 nonpolarized beam splitters. W is a $\lambda/2$ wave plate to control the polarization. RG is Rotating ground glass to randomize the phases of photons passing through it. M$_1$ and M$_2$ are mirrors. L$_1$ and L$_2$ are two identical lens with focus length of 100 mm and the distance between them are 200 mm. Acoustooptic modulator (AOM) is at the confocal point of L$_1$ and L$_2$ to change the frequency of laser light. H is a pinhole to block the laser light that does not change frequency after passing through AOM. L$_3$ and L$_4$ are two identical lens with focus length of 50 mm. S$_a$ and S$_b$ are point pseudothermal and laser light sources, respectively. FBS is a 1:1 nonpolarized fiber beam splitter. The distance between L$_3$ and the collector of FBS is equal to the distance between L$_4$ and the collector of FBS via BS$_2$. The optical length between the laser and detector via M$_1$ is 4.24 m. The single-photon counting rates of D$_1$ and D$_2$ are both about 50000 c/s, which means on average there is only $1.41 \times 10^{-3}$ photon in the experimental setup at one time. Our experiments are done at single photon's level.

There are four different cases to trigger a two-photon coincidence count in Fig. \ref{setup}. The first one is both photons are emitted by S$_a$. The second one is both photons are emitted by S$_b$. The third one is photon A is emitted by S$_a$ and photon B is emitted by S$_b$. The fourth one is photon A is emitted by S$_b$ and photon B is emitted by S$_a$. In the first case, there are two different alternatives to trigger a two-photon coincidence count, which are $A\rightarrow D_1, B\rightarrow D_2$ and $A\rightarrow D_2, B\rightarrow D_1$, respectively. In the second case, both photons are emitted by laser source S$_b$. There should be two alternatives, too. However, there is only one alternative since these two alternatives are identical (the reason why there is only one alternative will be discussed later). In the third case, there are two alternatives, which are $A\rightarrow D_1, B\rightarrow D_2$ and $A\rightarrow D_2, B\rightarrow D_1$, respectively. The fourth case is similar as the third one. For simplicity, the intensities of light beams emitted by S$_a$ and S$_b$ are assumed to be equal. If all the seven different alternatives are indistinguishable, the two-photon probability distribution function is \cite{feynman-p,feynman-l}
\begin{eqnarray}\label{G2-1}
&&G^{(2)}(\vec{r}_1,t_1;\vec{r}_2,t_2)\nonumber\\
&=& \langle |e^{i\varphi_{aA}}K_{aA1}e^{i(\varphi_{aB}+\frac{\pi}{2})}K_{aB2}+e^{i(\varphi_{aA}+\frac{\pi}{2})}K_{aA2}e^{i\varphi_{aB}}K_{aB1} \nonumber \\
&+& e^{i\varphi_{b}}K_{b1}e^{i(\varphi_{b}+\frac{\pi}{2})}K_{b2} \nonumber\\
&+& e^{i\varphi_{aA}}K_{aA1}e^{i(\varphi_{bB}+\frac{\pi}{2})}K_{bB2}+e^{i\varphi_{bB}}K_{bB1}e^{i(\varphi_{aA}+\frac{\pi}{2})}K_{aA2} \nonumber \\
&+& e^{i\varphi_{bA}}K_{bA1}e^{i(\varphi_{aB}+\frac{\pi}{2})}K_{aB2}+e^{i\varphi_{aB}}K_{aB1}e^{(i\varphi_{bA}+\frac{\pi}{2})}K_{bA2}|^2 \rangle.
\end{eqnarray}
Where $\varphi_{\alpha\beta}$ is the initial phase of photon $\beta$ emitted by source $\alpha$. $K_{\alpha\beta\gamma}$ is the probability amplitude that photon $\beta$ emitted by source $\alpha$ goes to detector $\gamma$ ($\alpha=a$ and b, $\beta=A$ and B, $\gamma=1$ and 2). The extra phase $\pi/2$ is due to the photon reflected by the fiber beam splitter will gain an extra phase comparing to the transmitted one \cite{loudon}.  $\langle...\rangle$ is ensemble average. The four lines on the righthand side of Eq. (\ref{G2-1}) correspond to the four different cases above, respectively.  Since the photons emitted by S$_a$ and S$_b$ are independent, $\langle e^{i(\varphi_a-\varphi_b)}\rangle$ equals 0. Taking the phase relationship into consideration, Eq. (\ref{G2-1}) can be simplified as
\begin{eqnarray}\label{G2-2}
&&G^{(2)}(\vec{r}_1,t_1;\vec{r}_2,t_2)\nonumber\\
&=& \langle |K_{aA1}K_{aB2}+K_{aA2}K_{aB1}|^2 \rangle \nonumber \\
&+& \langle|K_{b1}K_{b2}|^2\rangle \nonumber\\
&+& \langle|K_{aA1}K_{bB2}+K_{bB1}K_{aA2}|^2\rangle \nonumber \\
&+& \langle|K_{bA1}K_{aB2}+K_{aB1}K_{bA2}|^2 \rangle.
\end{eqnarray}
For a point light source, Feynman's photon propagator is \cite{peskin}
\begin{equation}\label{green}
K_{\alpha\beta\gamma}=\frac{\exp[-i(\vec{k}_{\alpha\gamma}\cdot\vec{r}_{\alpha\gamma}-\omega_{\alpha\gamma} t_{\alpha\gamma})]}{r_{\alpha\gamma}},
\end{equation}
which is the same as Green function in classical optics \cite{born}. $\vec{k}_{\alpha\gamma}$ and $\vec{r}_{\alpha\gamma}$ are the wave and position vectors of the photon emitted by S$_\alpha$ and detected at D$_\gamma$, respectively. $r_{\alpha\gamma}=|\mathbf{r}_{\alpha\gamma}|$ is the distance between S$_\alpha$ and D$_\gamma$. $\omega_{\alpha\gamma}$ and $t_{\alpha\gamma}$ are the frequency and time for the photon that is emitted by S$_\alpha$ and detected at D$_\gamma$, respectively ($\alpha=a$ and b, $\gamma=1$ and 2).

Substituting Eq. (\ref{green}) into Eq. (\ref{G2-2}) and with similar calculations in Refs. \cite{liu-2010,liu-2013,liu-2014-OC,liu-2014,shih-book}, it is straight forward to have one-dimension temporal two-photon probability distribution as
\begin{eqnarray}\label{G2-3}
&&G^{(2)}(t_1-t_2)\nonumber\\
&\propto &7+\text{sinc}^2\frac{\Delta\omega_a(t_1-t_2)}{2}\nonumber\\
&&+4\cos[\Delta\omega_{ab}(t_1-t_2)]\text{sinc}\frac{\Delta\omega_a(t_1-t_2)}{2}.
\end{eqnarray}
Where paraxial and quasi-monochromatic approximations have been employed to simplify the calculations. The positions of D$_1$ and D$_2$ are the same in order to concentrate on the temporal part. $\Delta\omega_a$ is the frequency bandwidth of pseudothermal light. $\Delta\omega_{ab}$ is the difference between the mean frequencies of pseudothermal and laser light. The first term on the righthand side of Eq. (\ref{G2-3}) is background coming from all four lines on the righthand side of Eq. (\ref{G2-2}). The second term on the righthand side of Eq. (\ref{G2-3}) is two-photon bunching of pseudothermal light coming from the first line on the righthand side of Eq. (\ref{G2-2}). The last term of Eq. (\ref{G2-3}) is two-photon interference which corresponds to the last two lines of Eq. (\ref{G2-2}). When the mean frequencies of these two superposed light beams are different, two-photon beating can be observed. The beating can only be observed within the second-order coherence time of pseudothermal light. The second-order coherence time is measured to be 51 $\mu$s, which is much larger than the beating period ($\approx 5 $ ns). The maximum visibility of the second-order interference pattern in Eq. (\ref{G2-3}) is 50\%.

\begin{figure}[htb]
    \centering
    \includegraphics[width=90mm]{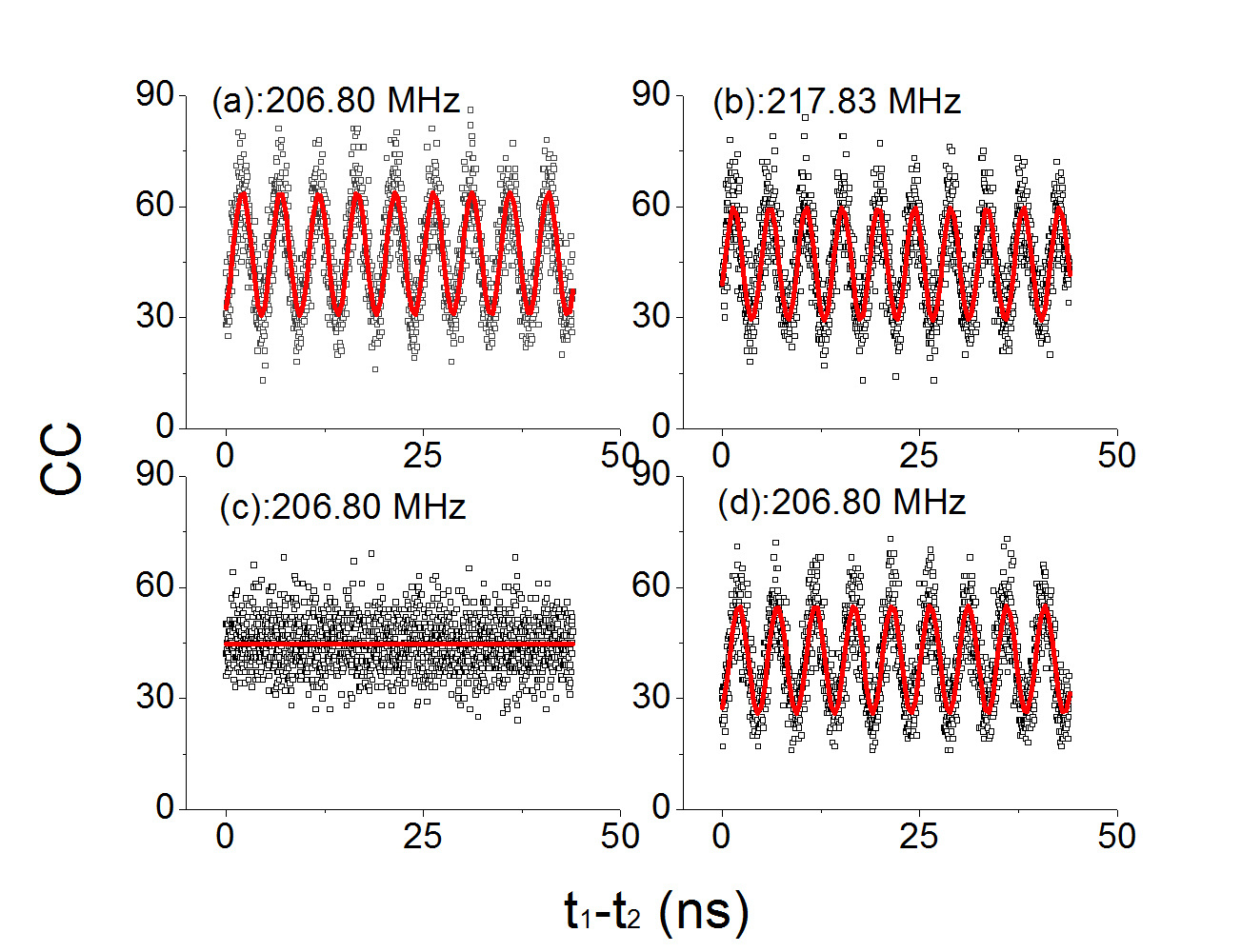}
    \caption{(Color online) Temporal second-order interference patterns. CC is two-photon coincidence counts for 600 s. $t_1-t_2$ is the time difference between the photon detection events of D$_1$ and D$_2$. The black squares are raw data without subtracting any background and red lines are sine fittings. See text for detail. }\label{beating}
\end{figure}

The measured temporal two-photon interference patterns are shown in Fig. \ref{beating}. CC is two-photon coincidence counts for 600 s. $t_1-t_2$ is the time difference between the photon detection events of D$_1$ and D$_2$. The black squares are raw data without subtracting any background and red lines are sine fittings. Figures \ref{beating}(a), (c) and (d) are taken when the frequency shift of AOM is 206.80 MHz and Fig. \ref{beating}(d) is taken when the frequency shift is 217.83 MHz. Two-photon beatings are observed in Figs. \ref{beating}(a) and (b) when the polarizations of the superposed light beams are parallel. The calculated beating frequencies are $206.02 \pm 0.08$ MHz and $218.15 \pm 0.07$ MHz, respectively. The calculated and measured frequency shifts are consistent. The reason why there is two-photon interference with non-identical photons is due to our detection system can not distinguish photons with frequency difference less than 2.2 GHz \cite{liu-2014,shih-book}. When the polarizations of the light beams emitted by these two sources are orthogonal and P$_2$ is removed, the photons are distinguishable by their polarizations. Two-photon beating disappears as shown in Fig. \ref{beating}(c). When we put P$_2$ back and set the axis at 45$^\circ$, the orthogonally polarized photons become indistinguishable. Two-photon beating is observed again in Fig. \ref{beating}(d). No matter the polarizations of the photons emitted by these two sources are parallel or orthogonal, the single photon counting rates of D$_1$ and D$_2$ are constant, which means the first-order interference pattern can not be observed by either D$_1$ or D$_2$. Although the photons are originated from the same single-mode laser, the initial phases and frequencies of photons emitted by S$_a$ and S$_b$ are different. Our experiments can be repeated with independent laser and thermal light beams as long as the frequency of each light beam is fixed during the measurement \cite{liu-2014}.

In the same way, we can observe two-photon interference with photons of different colors by employing a two-photon detection system with very short time measurement uncertainty. Let us take photons at 532 nm and 633 nm for example. The frequency difference between them is $5.65\times 10^{5}$ GHz, which is much larger than 2.2 GHz. Photons at 532 nm and 633 nm are distinguishable for our detection system and there is no two-photon interference. The time measurement uncertainty of two-photon absorption is at femtosecond range \cite{boitier-2009}, photons with frequency difference less than $ 10^{6}$ GHz are indistinguishable for the detection system. With two-photon absorption as the two-photon detection system, one may be able to observe two-photon interference with photons of different colors in the scheme of Fig. \ref{two}.

When single-photon sources instead of classical light sources are employed in Fig. \ref{two}, there are only two possible alternatives to trigger a two-photon coincidence count, which is given by the third or fourth line on the righthand side of Eq. (\ref{G2-2}). If there is a way to eliminate all the terms in Eq. (\ref{G2-2}) except the third and fourth lines, the observed two-photon interference pattern with single-mode laser would be the same as the one with single-photon sources. The visibility of two-photon interference with classical light can not exceed 50\% \cite{mandel-1983}. If two-photon absorption that only responds to $2\omega_{a0}+\Delta\omega$ is employed as two-photon detection system, only the last two lines of Eq. (\ref{G2-2}) are left for two-photon probability distribution function, where $\omega_{a0}$ and $\omega_{a0}+\Delta \omega$ are the central frequencies of photons emitted by S$_a$ and S$_b$, respectively. The visibility of two-photon interference pattern will exceed 50\% \cite{kaltenbaek-2009}. It seems strange that ``classical interference'' can be transformed into ``quantum interference'' with only the change of the detection system. As pointed out by Dirac \cite{dirac}, the superposition principles in quantum and classical physics are indeed different. The most important difference is that in classical physics where superposition principle holds, a state superposes with itself will get a different state with two times amplitude. However, a state superposes with itself in quantum physics will get the same state. It is the reason why two identical alternatives only contribute one alternative in Eq. (\ref{G2-1}) for photons in single-mode laser light. The difference between the superposition principles in quantum and classical physics is obviously different from the one between the quantum and classical interference discussed above.

In conclusions, whether there is two-photon interference or not for non-identical photons is dependent on the properties of photons and detection system, which can be analyzed by Heisenberg's uncertainty principle. Two-photon beating is observed when photons of different wavelengths are incident to the two input ports of a HOM interferometer, respectively. Although it is done with photons originated from the same laser, our experiment can be repeated with independent laser and thermal light sources. An experimental scheme is suggested to observe two-photon interference when photons of different colors are incident to two input ports of a HOM interferometer, respectively. The reason why two-photon interference phenomena are different for classical and nonclassical light is not interference, but the different properties of classical and nonclassical light and detection system. The studies clearly show that the observed results in quantum physics are dependent on the measurement apparatus. The conclusions can be generalized to any order of interference with photons and massive particles.

\section*{Acknowledgement}
The authors wish to thank D. Wei for the help on the AOM. This project is supported by National Science Foundation of China (No.11404255), Doctoral Fund of Ministry of Education of China (No.20130201120013), the 111 Project of China (No.B14040) and the Fundamental Research Funds for the Central Universities.

\end{document}